\documentclass[superscriptaddress,aps,showpacs,nofootinbib,twocolumn]{revtex4}

\usepackage{graphicx}
\usepackage{epsfig}
\usepackage{dcolumn}
\usepackage{amsmath}

\begin{document}
\title[BEC]{Squeezing and entanglement in quasiparticle\\
  excitations of trapped Bose-Einstein condensates\\} \author{J.
  Rogel-Salazar} \affiliation{Quantum Optics \& Laser Science,
  Department of Physics, Imperial College, London SW7 2BW, U.K.}
\author{S. Choi} \affiliation{Clarendon Laboratory, Department of
  Physics, University of Oxford, Parks Road, Oxford OX1 3PU, U.K.}
\author{G.H.C. New} \affiliation{Quantum Optics \& Laser Science,
  Department of Physics, Imperial College, London SW7 2BW, U.K.}
\author{K. Burnett} \affiliation{Clarendon Laboratory, Department of
  Physics, University of Oxford, Parks Road, Oxford OX1 3PU, U.K.}

\begin{abstract}
  \vspace{0.3in}We estimate the amount of temperature-dependent
  squeezing and entanglement in the collective excitations of trapped
  Bose-Einstein condensates.  We also demonstrate an alternative
  method of temperature measurement for temperatures much less than
  the critical temperature $(T\ll T_c)$.
\end{abstract}
 \maketitle

\section{Introduction}
\label{sec:introduction}

Since the first experimental realization of Bose-Einstein condensation
in trapped atomic gases, a number of papers have highlighted the
parallels between processes in atom optics and nonlinear (photon)
optics.  Well-known examples include four-wave mixing \cite{4wm} and
soliton dynamics \cite{reinhardt}, while the squeezing of matter wave
fields has also been examined \cite{duan,sorensen,EPR}.  These latter
studies consider the way in which the nonlinear interactions between
the atoms can generate entangled atomic beams, for example by
spin-exchanging collisions of spinor Bose-Einstein condensates.

Entangled states, to which squeezed states are closely related, have
been extensively studied in the last few years because of their
potential application in quantum information processing and quantum
computing. In this paper, we discuss squeezing and entanglement in
terms of the quasiparticle modes of a trapped gas.  Quasiparticle
modes represent collective excitations of the Bose-Einstein condensate
(BEC), and are in fact one of the most fundamental features of its
dynamics.  They are easily generated by applying time-dependent
perturbations to the trapping potential, and were observed
experimentally quite early on; they have since been studied in some
detail both theoretically and experimentally.

The low temperature regime in which the majority of the atoms are in
the condensate has been studied in various experiments
\cite{jin1,mewes}. These experiments revealed oscillations of the
condensate with almost no damping, and the results were in good
agreement with predictions based on the zero temperature
Gross-Pitaevskii equation (GPE) \cite{edwards}.  Excitations at higher
temperatures were studied in later experiments \cite{jin2}, where
large energy shifts and rapid damping rates were observed.  While the
standard theory of elementary excitations due to Bogoliubov
\cite{bogo} explains most of the observed effects, it cannot account
for higher-order processes, such as the Beliaev damping. And while
finite temperature studies using the GPE have explained various
properties of BECs observed experimentally, they cannot fully describe
the evolution of collective excitations at higher temperatures
\cite{dodd} since the spontaneous part of the Beliaev damping of the
excitations is not included.  Theoretical techniques have recently
been developed that take the quasiparticle interactions into account
\cite{sam} and experiments to test the predictions are planned for the
near future. The main motivation of the present paper is to analyse
these experiments on the quantal aspects of quasiparticle mode
evolution in a BEC.

We note that Beliaev damping may be understood in terms of a nonlinear
frequency-mixing mechanism in which a Bogoliubov quasiparticle of
frequency $\omega_2$ interacts with the ground state (condensate) and
generates two quasiparticles of frequency $\omega_1$ that divide the
initial energy equally (Fig. \ref{fig:diag}). The mechanism is
actually based on a four-wave process that masquerades as a three-wave
interaction analogous to optical parametric down-conversion (OPDC) in
which a photon in a nonlinear medium splits into two photons of lower
energy.

We calculate the amount of squeezing for a realistic system containing
10000 $^{87}$Rb atoms in a trap with spherical geometry. We also
propose an alternative method for determining the temperature of a BEC
well below the critical temperature by observing the variation of the
envelope of the collective excitations. This should be more accurate
than current techniques based on fitting a thermal Gaussian profile to
the atomic cloud.

The paper is organised as follows: In Section \ref{sec:hamiltonian},
we describe the higher-order Hamiltonian that accommodates processes
such as Landau and Beliaev damping in the quasiparticle excitations.
We derive equations of motion for a quasiparticle mode.  In Section
\ref{sec:pdc}, we make quantitative estimates for the degree of
squeezing and entanglement in quasiparticle excitations.  In addition
we calculate the expected ``damping rate'' and coupling between modes
at various temperatures which can then serve as a temperature
measurement calibration for ultracold atoms. A possible way of
detecting squeezed states in atom optics is discussed briefly in
Section \ref{sec:experiment}.

\section{Hamiltonian in the quasiparticle basis}
\label{sec:hamiltonian}

The many-body Hamiltonian for a system of bosons with pairwise
interactions can be written in the usual second quantised formalism
as,
\begin{equation}
  \label{eq:bareH}
  \hat H= \sum_{ij} H_{ij}^{sp} \hat a^{\dag}_i \hat a_j+\frac{1}{2}
  \sum_{ijkm} \langle ij|\hat V|km\rangle \hat a_{i}^{\dag}
  \hat a_{j}^{\dag} \hat a_{k} \hat a_{m},
\end{equation}
where the matrix elements $H{ij}^{sp}$ are given by
\begin{equation}
  \label{eq:Hsp}
  H_{ij}^{sp}=\int d^3 {\bf r} \psi_i^*({\bf r})\hat H^{sp}\psi_j({\bf r}).
\end{equation}

Here, $H^{sp}=-\frac{\hbar}{2m}\nabla^2+V_{trap}$ is the
single-particle Hamiltonian with a confining potential $V_{trap}$, and
the basis state wave functions are $\psi_i ({\bf r})$. $\langle
ij|\hat V|km\rangle$ denotes the matrix element for the interaction
potential $\hat V(\mathbf{r})$ between atoms. The operators $\hat
a_{i}^{\dag}$ and $\hat a_{i}$ are the creation and annihilation
operators for mode $i$ that obey the usual Bose commutation relations
\begin{equation}
  [\hat a_i,\hat a_j^{\dag}]=\delta_{ij},\text{\ \ \ \ \ }
  [\hat a_i,\hat a_j]=[\hat a_i^{\dag},\hat a_j^{\dag}]=0.\nonumber
\end{equation}

The Hamiltonian (\ref{eq:bareH}) is written in a single-particle basis
where the operator $\hat a_i$ annihilates a particle from the state
with wave function $\psi_i({\bf r})$. The wave function $\psi_0({\bf
  r})$ describes the condensate, while the remaining functions form a
complete set orthogonal to the condensate.

We follow the standard procedure and rewrite the Hamiltonian
(\ref{eq:bareH}) in powers of $\hat{a}_i$ by replacing the operator
$\hat a_0$ with $\sqrt{N_0}=(N-N_{ex})^{1/2}$ where $N$ is the total
number of atoms in the trap and $N_{ex}=\sum_{i \neq 0} \hat
a_i^{\dag} \hat a_i$ is the number of excited atoms. The quadratic
part of the Hamiltonian can be diagonalised exactly using the standard
Bogoliubov transformation \cite{sam}, which allows us to write the
Hamiltonian in a quasiparticle basis. The quasiparticle operators
$\hat \beta_i$ are defined by\footnote{For simplicity from now on we
  drop the explicit notation in the operators}
\begin{equation}
  \label{eq:quasi}
  \beta_i=\sum_{j \neq 0}u_{ij}^* a_j-v_{ij}^* a_j^{\dag},
\end{equation}
where the matrices $u_{ij}$ and $v_{ij}$, which depend on temperature
and the trap geometry, are calculated numerically using the
Bogoliubov-de Gennes (BdG) formalism \cite{hutchinson}.

The non-quadratic terms, which affect the energy and shape of the
condensate, are expected to be small and hence dealt with
perturbatively. We can, therefore, calculate the quasiparticle shifts
and widths at zero and finite temperature. It has been shown
\cite{sam} that the effective Hamiltonian can be written in terms of
the quasiparticle operator $\beta_i$ as
\begin{widetext}
\begin{equation}
  H'= const + \sum_{i \neq 0} (\epsilon_i+\Delta \epsilon_i)\beta_i^{\dag}
  \beta_i+\left\{\sum_{ijk \neq 0} \left [\zeta_{ijk} \beta_i^{\dag} \beta_j \beta_k\right]+h.c.
  \right\},
  \label{eq:H3}
\end{equation}
\end{widetext}
in which only processes that generally conserve energy have been
included. The constant term in the equation simply defines the zero of
energy and $\Delta \epsilon_i$ is the energy shift from first order
perturbation theory. The coefficients $\zeta_{ijk}$ are determined by
\begin{widetext}
\begin{eqnarray}
  \zeta_{ijk} &=& \sqrt{N_0} \sum_{mnq \neq 0} \langle q0|V|mn \rangle
  \left [u_{iq}^* u_{jn} u_{km}+v_{in}^* v_{jq} u_{km}+v_{im}^* u_{jn}
  v_{kq}\right ]\nonumber\\
  & & + \langle mn|V|q0 \rangle \left [u_{in}^* u_{jq} v_{km}+u_{im}^*
  v_{jn} u_{kq}+v_{iq}^* v_{jn} v_{km} \right ].
\end{eqnarray}
\end{widetext}
where the indices $i,j$ and $k$ are labels that denote the
quasiparticle energy levels. For a 3-D condensate, the index $i$
stands for the quantum numbers $n,l$ and $m$. We note that the
coefficients are temperature dependent through $u_{ij}$ and $v_{ij}$.

The Hamiltonian given by equation (\ref{eq:H3}) contains terms beyond
the standard Bogoliubov approximation, thus we are using a fuller
description of the BEC that involves {\it interacting} quasiparticles
and can take important processes such as Landau and Beliaev damping
into account. Beliaev processes occur at zero temperature and are
dominant in the low-temperature regime. Landau processes on the other
hand predominate at higher temperatures; these processes, in which two
quasiparticles collide to form a single quasiparticle cannot occur at
zero temperature because there are no excited quasiparticles.
 
From the Hamiltonian (\ref{eq:H3}), the Heisenberg equations of motion
for $\beta_p$ and $\beta_p^\dag$ are
\begin{widetext}
\begin{eqnarray}
  \label{eq:beta}
  {\rm i} \dot \beta_p & = & \omega_p\beta_p+
  \sum_{j,k \neq 0}\sigma_{jk}\beta_j \beta_k+\sum_{j,k \neq 0}
  \nu_{jk}\beta_k^\dag \beta_j,\\
  \label{eq:betadag}
  - {\rm i} \dot \beta_p^\dag & = & \omega_p\beta_p^\dag+
  \sum_{j,k \neq 0}\sigma_{jk}^*\beta_j^\dag
  \beta_k^\dag+\sum_{j,k \neq 0} \nu_{jk}^*\beta_j^\dag \beta_k,
\end{eqnarray}
\end{widetext}
where
\begin{eqnarray*}
\omega_p&=&\left(\epsilon_p+\Delta \epsilon_p\right)/\hbar,\\
\sigma_{jk}&=&\zeta_{pjk}/\hbar,\\
\nu_{jk}&=&\left(\zeta_{jpk}^*+\zeta_{jkp}^*\right)/\hbar.
\end{eqnarray*}

It is well known that homogeneous systems have a continuous spectrum
of excitations; however, for a trapped system, the spectrum of states
to which the excitation can couple is discrete.  Furthermore, in the
case of a spherical trap there is a degeneracy in the azimuthal
quantum number $m$, which in turn means that fewer resonances are
present \cite{guilleumas,martin}. This implies that the time evolution
of the excitation is dominated by a strong coupling to only a few
modes. On the other hand, the geometry of the trap can be changed by
adjusting the frequencies in the radial and axial directions
independently, and the selection of a dominant mode in this way has
been demonstrated \cite{hechen}. In particular, a Beliaev process has
recently been observed for a scissors mode, where one mode is
resonantly coupled to two modes of half the original frequency
\cite{beldamp}; this is the kind of process that we seek to model in
the present paper. The equations of motion for modes $p=1,2$ in a
Beliaev process reduce to
\begin{eqnarray}
  \label{eq:twomodes1}
  \dot \beta_1 & = & -{\rm i}\omega_1\beta_1-{\rm i}\nu_{21}
  \beta_2\beta_1^{\dag},\\
  \dot \beta_1^\dag & = & {\rm i}\omega_1\beta_1^\dag+{\rm i}\nu_{21}
  \beta_2^{\dag}\beta_1,\\
  \label{eq:twomodes}
  \dot \beta_2 & = & -{\rm i}\omega_2\beta_2-{\rm i}\frac{\nu_{21}}{2}
  \beta_1\beta_1,\\
  \label{eq:twomodes2}
  \dot \beta_2^\dag & = & {\rm i}\omega_2\beta_2^\dag+{\rm i}
  \frac{\nu_{21}}{2}\beta_1^{\dag}\beta_1^{\dag}.
\end{eqnarray}
where we have used the fact that $\sigma_{11}=\nu_{21}/2$, and we have
chosen the phases so that the coefficients are real.

Equations of the form (\ref{eq:twomodes1})-(\ref{eq:twomodes2}) have
been studied in quantum optics and we apply them here to study the
squeezing in the quasiparticle excitations and its temperature
dependence. We note that damping of quasiparticle excitations has
earlier been described in terms of nonlinear mixing of quasiparticle
modes \cite{NLM}; however, the present work is quite distinct as the
operator nature of the quasiparticle annihilation operator $\beta_i$
is retained, and the calculation is not restricted to the quadratic
approximation. We are therefore extending the work of Ref.
\cite{NLM}, so that spontaneous quantum processes are included.

To give an idea of the order of magnitude of the numbers involved, the
coefficient $\nu_{21}$ equals $1.9\times 10^{-2}$ in trap units for
the following parameters: $T=20$ nK, $N=10000$,
$\omega_{trap}=2\pi\times 100$ Hz, for a spherical trap geometry and
using $^{87}$Rb atoms for which the s-wave scattering length $a=110
a_0$, where $a_0$ is the Bohr radius.  The values of $\nu_{21}$ are
significantly dependent on temperature; which implies that the
equations of motion for $\beta_1$ and $\beta_1^\dag$ are altered
accordingly. We can exploit this feature as a method of measuring
temperature below $T_c$.  Figure \ref{fig:TB} indicates the dependence
of $\nu_{21}$ on temperature, which is an effective dependence on the
number of particles in the condensate. We notice that at higher
temperatures the coefficient decreases swiftly, whilst for lower
temperatures it is fairly constant.

\section{Temperature dependent coupling process}
\label{sec:pdc}

\subsection{Non-depleted regime}
\label{sec:nondeplete}

We consider first the case where the higher ($p=2$) mode - the pump -
has a much larger population than the lower one and where the pump
depletion is ignored so that the operator $\beta_2$ can be
approximated by a c-number $b_2$. Physically, this would represent a
situation in which the higher mode is being continuously driven by a
resonant excitation. The solutions of the equations of motion for
$\beta_1$ and $\beta_1^{\dag}$ in the interaction picture are then
given by
\begin{equation}
  \label{eq:solution}
  \left( \begin{array}{c}
    {\beta_1(t)}\\ {\beta_1^{\dag}(t)}
  \end{array} \right) = \left( 
  \begin{array}{cc}
    \cosh(\Omega t) & -{\rm i}\sinh(\Omega t)\\ {\rm i}\sinh(\Omega t)
  & \cosh(\Omega t)
  \end{array} \right) \left(
  \begin{array}{c}
    \beta_1(0) \\ \beta_1^{\dag}(0)
  \end{array} \right),
\end{equation}
where $\Omega=\nu_{21}b_2$, and we have chosen the phases such that
the coefficients are real. This approximation neglects the depletion
of the ``pump'' and the solution will cease to be valid once
appreciable down-conversion occurs.

Solutions (\ref{eq:solution}) are formally equivalent to equations
that describe degenerate parametric down conversion in quantum optics:
a process known to be a significant source of squeezed states. In our
case the squeezing parameter is given by

\begin{equation}
  \label{eq:sqparameter}
  \tau=\Omega t,
\end{equation}
which is temperature dependent through $\Omega$.

Defining the quadrature operators as
$X(t)=\frac{\beta_1(t)+\beta_1^\dag(t)}{2}$ and
$Y(t)=\frac{\beta_1(t)-\beta_1^\dag(t)}{2{\rm i}}$, we have the
variances
\begin{eqnarray}
  \label{eq:x1}
  (\Delta X(t))^2&=&\frac{(2N_1 +1)}{4} \exp({-2\Omega t}),\\ 
  \label{eq:x2}
  (\Delta Y(t))^2&=&\frac{(2N_1 +1)}{4} \exp({2\Omega t}),
\end{eqnarray}  
where $N_1$ is the number of particles in mode 1. In optics, the
quadrature operators are well-defined quantities corresponding to the
amplitude and phase of the electromagnetic (EM) oscillation.  One may
think of the amplitude and (temporal) phase of oscillations in a
quasiparticle excitation, in a similar way, and we therefore interpret
our quadrature operators $X$ and $Y$ as amplitude and phase of
oscillations.  Equations (\ref{eq:x1}) and (\ref{eq:x2}) imply that
the degree of squeezing depends on the amount of the lower-mode atoms
present. This is directly related to the temperature of the ultracold
atoms in the condensate \cite{rich}, defined in terms of the initial
Bose-Einstein (BE) distribution of quasiparticles.  Surface plots of
$\Delta X$ and $\Delta Y$ are shown in Figures \ref{fig:varianzas} a)
and b). At $t=0$ the behaviour of the variances is determined by the
BE distribution function for mode $1$. At higher temperatures, the
behaviour of the coefficient $\nu_{21}$ dominates the trend.

Experimentally, what is of interest is correlation functions such as
$\langle \beta_1\beta_1 \rangle$ or $\langle \beta_1^\dag\beta_1
\rangle$, among others.  These quantities have a number of significant
physical interpretations.  Assuming an initial number state in the
quasiparticle basis, some important correlation functions are given
by:
\begin{eqnarray}
  \langle \beta_1(t) \beta_1(t) \rangle &=& -{\rm i}\left(N_1+
  \frac{1}{2}\right)\sinh(2\Omega t),\\
  \langle \beta_1^\dag(t) \beta_1(t) \rangle &=& \left(N_1+
  \frac{1}{2}\right)\cosh(2\Omega t)-\frac{1}{2}.\label{eq:N1}
\end{eqnarray}

A surface plot of $\langle \beta_1\beta_1 \rangle$ as a function of
temperature, $T$ and time, $t$ is given in Figure \ref{fig:betasb} a);
a non-zero value of this quantity implies the presence of squeezing
and hence entanglement. The correlation function of equation
(\ref{eq:N1}) is just the population in the lower mode, and is
directly measurable as the amplitude of oscillation. A surface plot of
this quantity is shown in Figure \ref{fig:betasb} b) as a function of
temperature $(T)$ and time $(t)$.  We note that at $t=0$ the trend of
both correlation functions is given by the BE distribution function
for the population in the lower mode. As time evolves, the population
of the lower mode increases. The solution is valid within the
non-depleted regime, where mode 2 is continually replenished.

\subsection{Depleted regime}
\label{sec:depleted}

We now examine the case where mode 2 is depleted and an explicit
analytical solution is possible. The analysis is handled more easily
by considering the equations of motion for the number operators
$N_i=\beta_i^{\dag}\beta_i$, which from equations (\ref{eq:twomodes1}
- \ref{eq:twomodes2}) are
\begin{eqnarray}
  \label{eq:N_1} 
  \frac{{\rm d}{N_1}}{{\rm d} t}&=&{\rm i}\nu_{21}
  \left(\beta_2^\dag\beta_1\beta_1-\beta_1^\dag\beta_1^\dag\beta_2\right),\\
  \label{eq:N_2}
  \frac{{\rm d}{N_2}}{{\rm d} t}&=&{\rm i}\frac{\nu_{21}}{2}\left(\beta_2
  \beta_1^\dag\beta_1^\dag-\beta_2^\dag\beta_1\beta_1\right).
\end{eqnarray}

Equations (\ref{eq:N_1}) and (\ref{eq:N_2}) can, in fact, be uncoupled
by calculating the second derivative and using the fact that the
operator $A=N_1+2N_2$ is a constant of motion. The uncoupled equations
are given by
\begin{eqnarray}
   \label{eq:uncoupN1}
   \frac{{\rm d}^2{N_1}}{{\rm d} t^2}&=&\nu_{21}^2(-3N_1^2+2AN_1+A),\\
   \label{eq:uncoupN2}
   \frac{{\rm d}^2{N_2}}{{\rm d} t^2}&=&\frac{\nu_{21}^2}{2}(12N_2^2-8AN_2+A^2-A).
\end{eqnarray}
In the case in which $N_2$ is strong, we can determine the population
by taking the average of equation (\ref{eq:uncoupN2}). We end up with
a c-number second-order differential equation with the following
initial conditions
\begin{eqnarray}
  \label{eq:dotN2} 
  \left.\frac{{\rm d}N_2}{{\rm d}t}\right|_{t=0}&=&0,\\ 
  \label{eq:N20} 
  N_2(0)&=&N_{20}; 
\end{eqnarray}
whose solution is given by
\begin{widetext}
\begin{equation}
\label{eq:sol1N2} 
N_2(t) = \left\{ 
\begin{array}{ccc} 
    N_{20}+(\alpha_2-N_{20}){\rm sn}^2\left
   (\frac{\nu_{21}}{2}\sqrt{\frac{\alpha_1-N_{20}}{6}}t
   ,\sqrt{\frac{\alpha_2-N_{20}}{\alpha_1-N_{20}}}\right)
   , \text{\ \  for } N_{20}<\alpha_2,\\
    \\
    \alpha_1+(N_{20}-\alpha_1){\rm nd}^2\left(\frac{\nu_{21}}{2}
    \sqrt{\frac{\alpha_1-\alpha_2}{6}}t,\sqrt{\frac{N_{20}-\alpha_2}
    {\alpha_1-\alpha_2}}\right), \text{\ \  for } N_{20}>\alpha_2.
\end{array}
\right. 
\end{equation} 
\end{widetext}
Here ${\rm sn}(u,k)$ and ${\rm nd}(u,k)$ are Jacobi elliptic functions
\cite{byrd,appell}, $\alpha_1$ and $\alpha_2$ are the roots of the
quadratic polynomial
$P(N_2)=-N_2^2+(A-N_{20})N_2-N_{20}^2-(A^2-A)/4+N_{20} A$.  As we have
pointed out, the coupling coefficient $\nu_{21}$ depends on
temperature; consequently the behaviour of $N_2$ must change with $T$.
The initial value $N_{20}$ was calculated for each temperature
according to the BE distribution function and an initial driving was
also considered.  These parameters were then used in the solution
(\ref{eq:sol1N2}).  Figure \ref{fig:population} is a plot of the
population $N_2$ showing its dependence on temperature and time.

In a practical case, the fitting of an experiment to the theory can be
realized by noticing that the initial part of the curve for the
population in the quasiparticle mode 2 could be approximated by a
function of the form $a_1\cos(\Gamma t)$ (Fig.  \ref{fig:graphn2}). To
a first approximation the quantity $\Gamma$ is given by
\begin{equation}
  \label{eq:gammanu}
  \Gamma=\nu_{21}\sqrt{\frac{(\alpha_1-N_{20})(N_{20}-\alpha_2)}{12 N_{20}}}.
\end{equation}

The full solution, given by equation (\ref{eq:sol1N2}), has been used
in a fitting routine and a plot of $\Gamma$ as a function of
temperature is shown in Figure \ref{fig:gamma}. In principle, $\Gamma$
has the advantage of being readily measurable. At lower temperatures
the effect of the initial population is dominant, while, for higher
temperatures, the coupling coefficient $\nu_{21}$ decreases rapidly so
that the effect of the initial population is masked, which happens at
approximately $0.6 T_c$. This effect is the result of having a
competition between the changing coupling strength and the influence
of the initial population.  It is noted that one could also consider
$\Gamma$ as an experimentally accessible indicator of the amount of
squeezing in the quasiparticles, as it is proportional to the coupling
strength $\nu_{21}$.

\section{Discussion} 
\label{sec:experiment} 

Squeezed states of the electromagnetic field were realized some time
ago and the degree of squeezing can be measured with standard
techniques such as homodyne detection. In the case of squeezed
quasiparticle modes of a BEC, it will be necessary to devise a method
for their detection. It is noted that a related work that uses neutron
scattering was suggested by Yurke \cite{yurke}.

We suggest for our system that Raman scattering between the two
excited phonon states could be used to characterise the amount of
squeezing present since single particle transitions will be strongly
modified by the presence of a correlated pair excitation. This effect
has been demonstrated for a homogeneous gas by Stamper-Kurn et.al.
\cite{stamper} who used a single particle transition to excite a
phonon. We then have two routes for the transfer of momentum. The
first moves a particle with zero momentum to momentum $k$, while the
second takes the momentum from $-k$ to zero. The interference between
the two routes will influence the single particle transition. Raman
scattering will be strongly influenced by the presence of coherent
states in the lowest mode. The precise form will depend on the
conditions, pulses used, and other experimental parameters
\cite{recwork}. Similar ideas have been used to probe coherent phonons
in solid state physics \cite{misochko}.

In summary, we have analysed a process analogous to parametric
down-conversion in trapped Bose-Einstein condensates. Physically this
corresponds to Beliaev damping, where two quanta are produced by one
of higher frequency. This, in turn, suggests the possibility of
observing temperature-dependent squeezing in the collective
excitations of the condensate. The amount of squeezing gives a direct
indication of the temperature and vice versa.

\begin{acknowledgments} 
  This work has been supported by CONACyT, EPSRC, the Royal Commission
  for the Exhibition of 1851 and the EU.
\end{acknowledgments}

\begin{center}
  \newpage {\Large Figure Captions}
\end{center}
\begin{enumerate}
\item \label{fig:diag} Schematic diagram of energy levels for a
  trapped Bose-Einstein condensate.  In the Beliaev process, analogous
  to optical parametric down-conversion, a quasiparticle of frequency
  $\omega_2$ interacts with the ground state generating two
  quasiparticles of frequency $\omega_1$ that divide the initial
  energy equally.
  
\item \label{fig:TB} Coefficient $\nu_{21}$ vs temperature for
  $N=10000$, $\omega_{trap}=2\pi\times 100$ Hz, a spherical trap
  geometry and using $^{87}$Rb atoms.
  
\item \label{fig:varianzas} a) Surface plot of $\Delta X$ as a
  function of temperature and time.  b) Surface plot of $\Delta Y$ as
  a function of temperature and time.  At $t=0$, the behaviour is
  fully determined by the Bose-Einstein (BE) distribution function.
  
\item \label{fig:betasb} Correlation functions calculated within the
  non-depleted regime: a) The average $\langle \beta_1\beta_1
  \rangle$. The occurrence of squeezing is implied by a non-zero value
  of this quantity. b) Surface plot of $\langle \beta_1^\dag\beta_1
  \rangle$, which describes the behaviour of the population in mode 1.
   
\item \label{fig:population} Surface plot of the population of mode 2
  as a function of time and temperature.
    
\item \label{fig:graphn2} Plot of $N_2(t)$, the solid line is the
  solution for the following parameters: $N_{20}=151.4225$, at
  $T=20nK$. The dashed line represents the amplitude of oscillations
  of the quasiparticle mode 2.
  
\item \label{fig:gamma} The parameter $\Gamma$ plotted against
  temperature is obtained so that the curve $a_1 \cos(\Gamma t)$
  corresponds to the envelope shown in figure \ref{fig:graphn2}.
    
\end{enumerate}

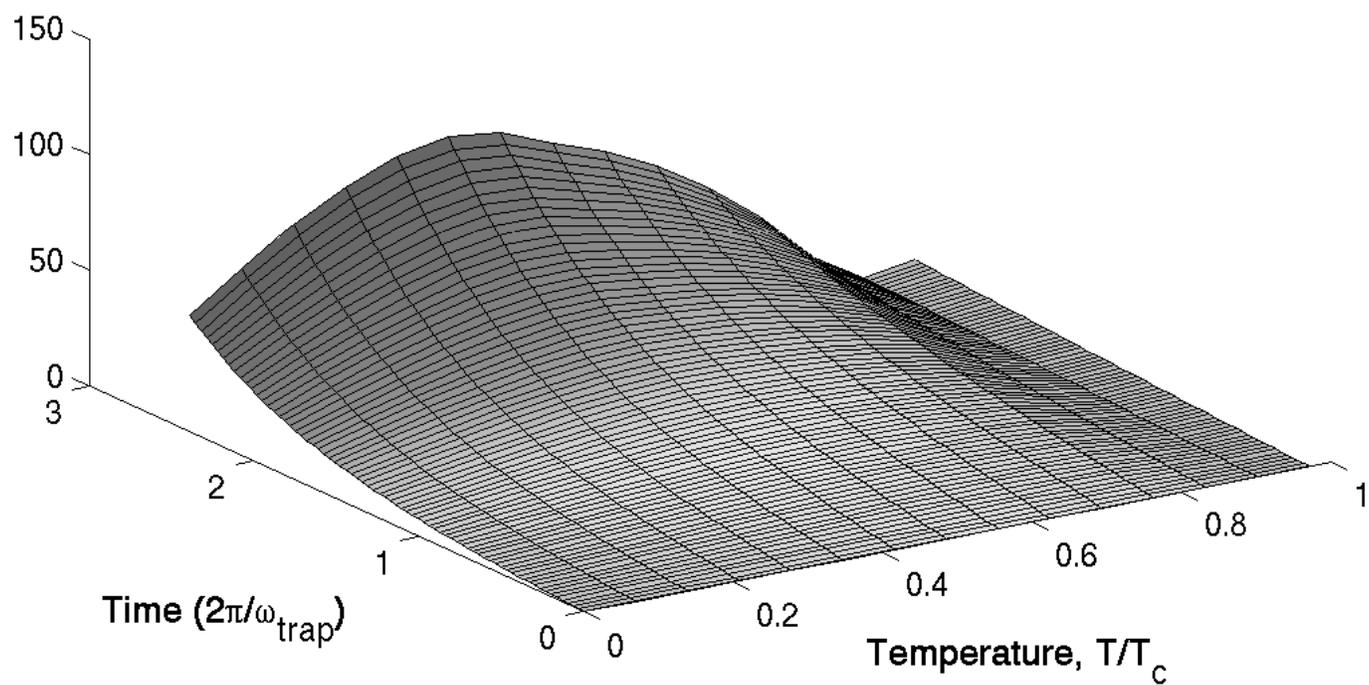
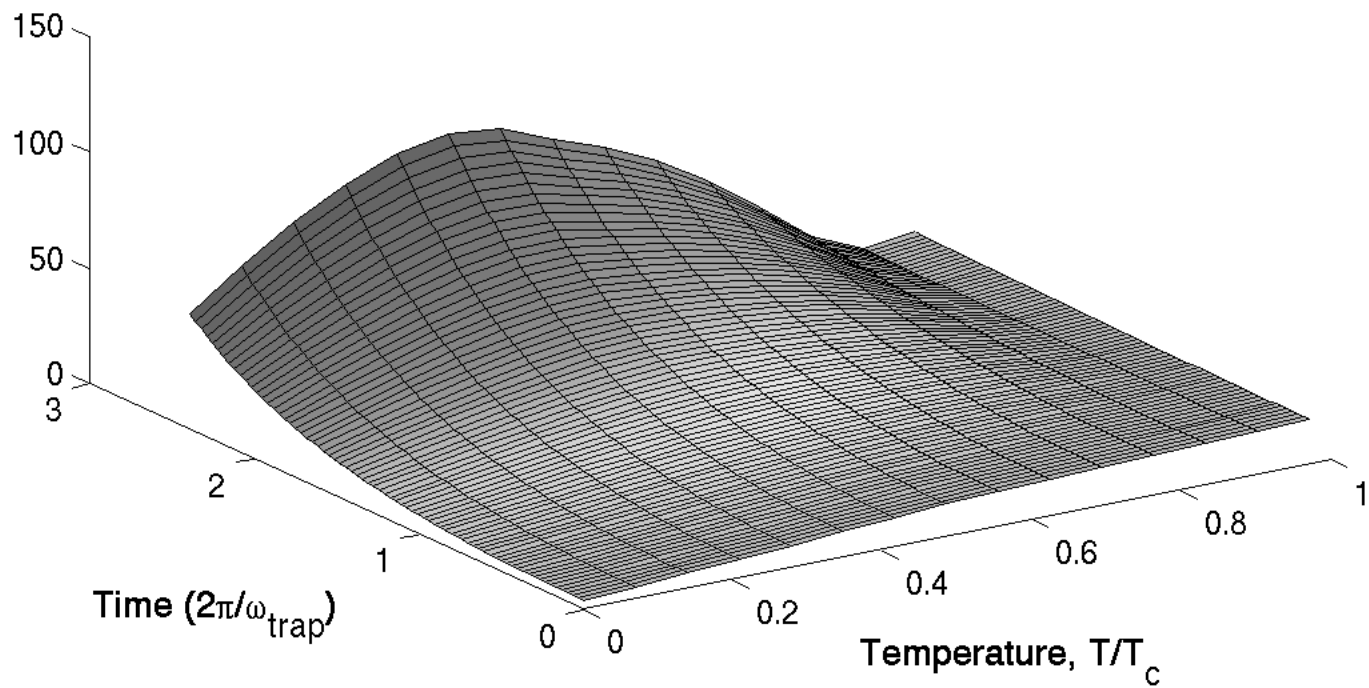
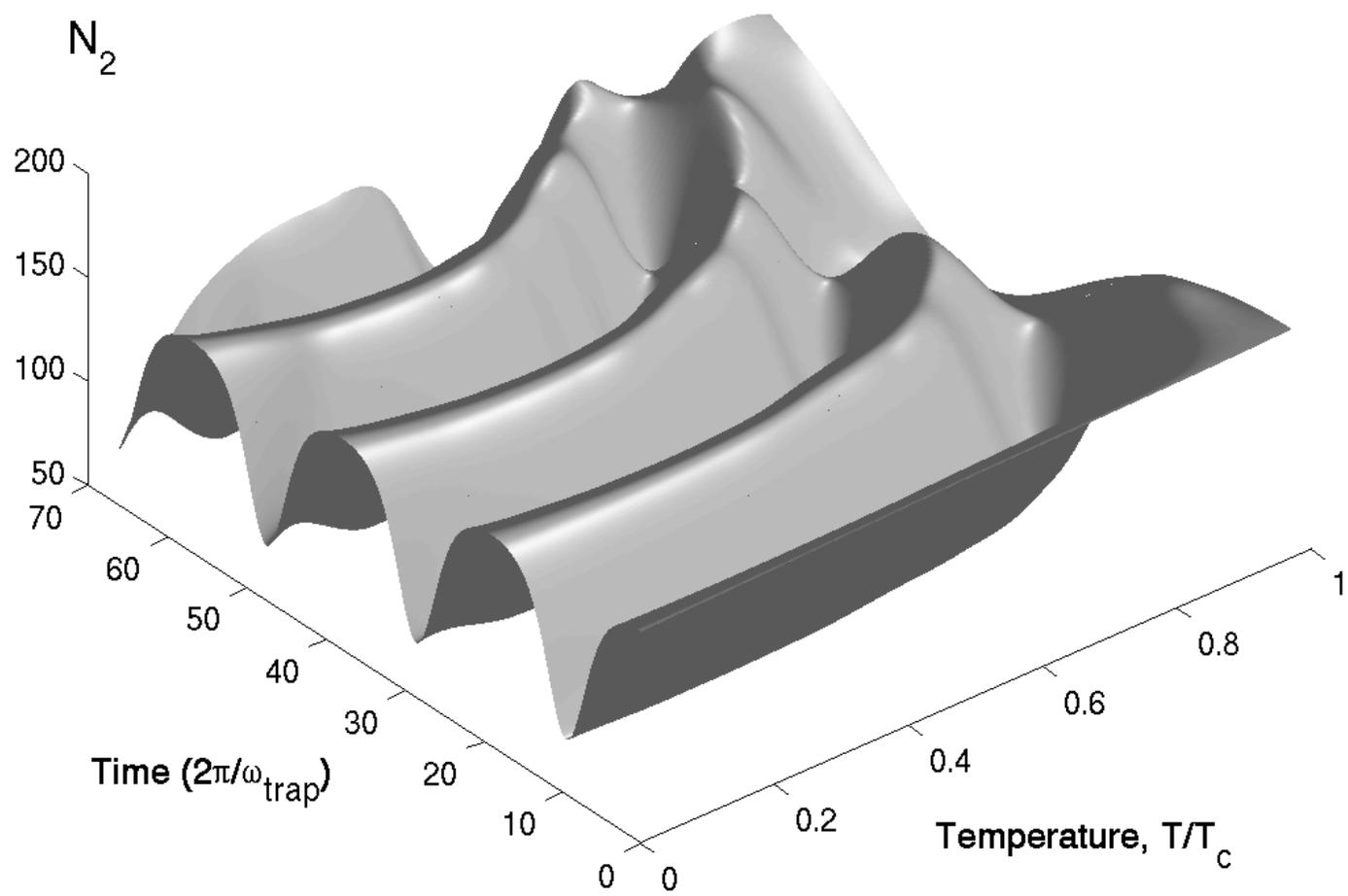
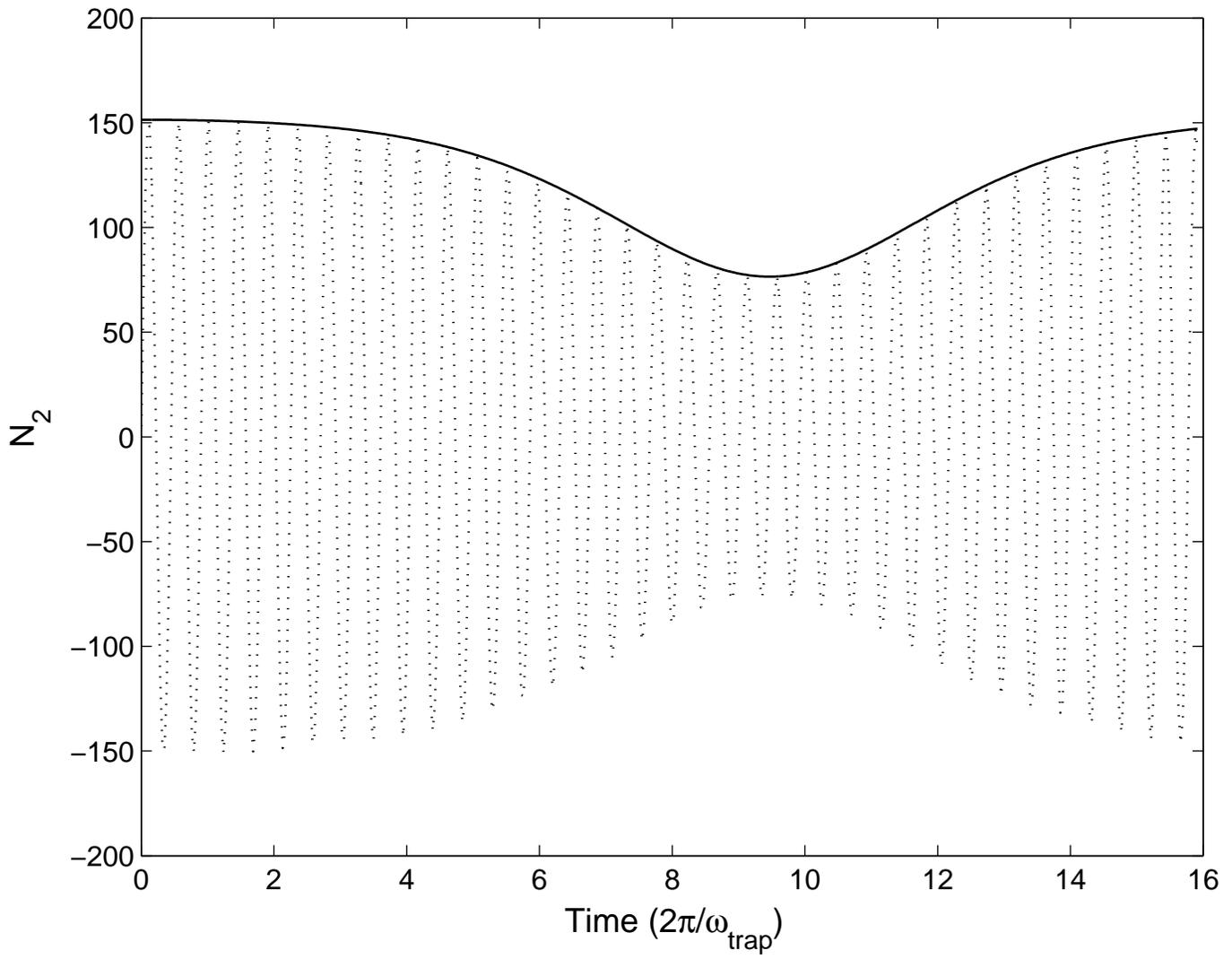
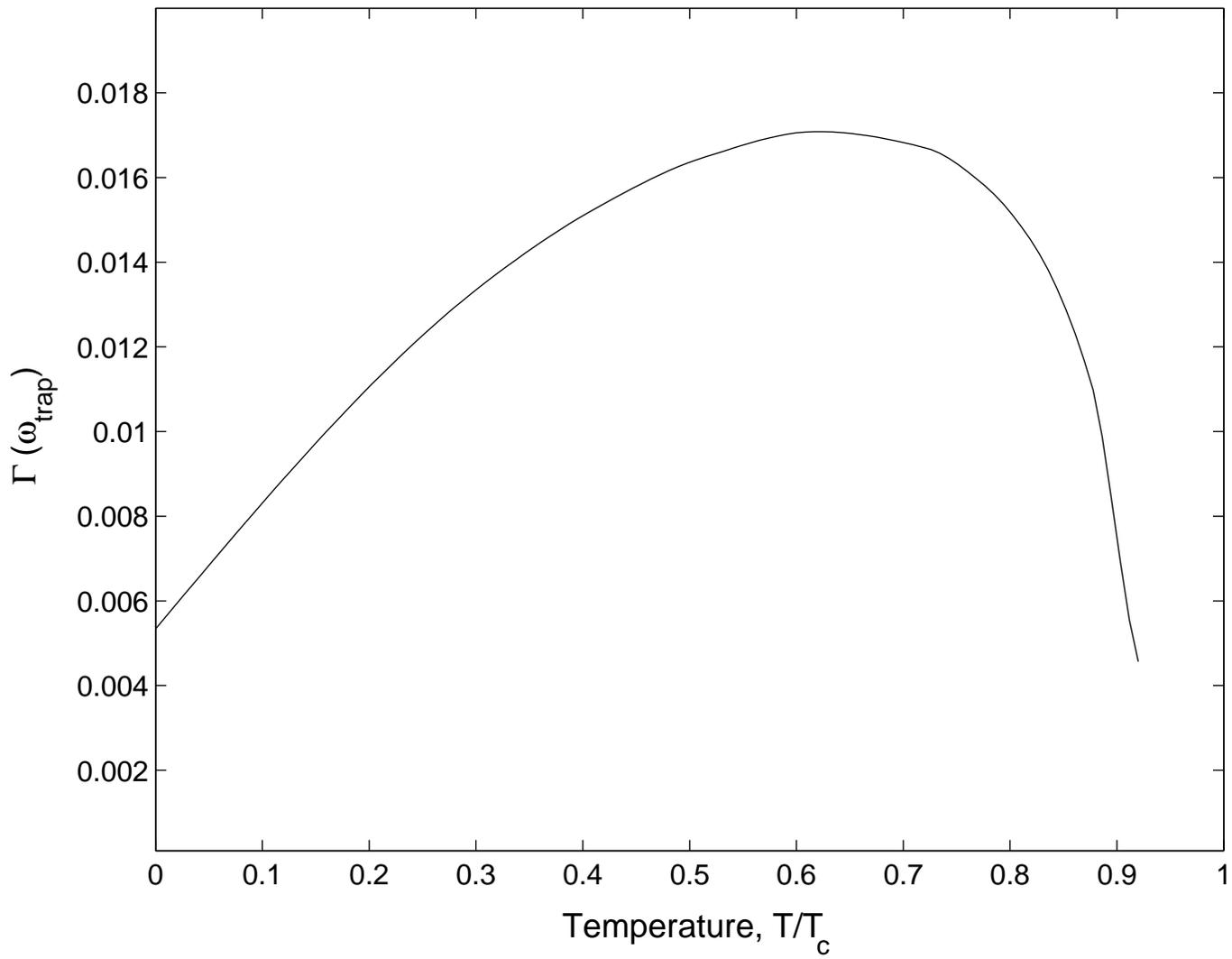
\end{document}